\documentclass[acmtog]{acmart}

\usepackage{multicol}
\usepackage[utf8]{inputenc}

\AtBeginDocument{%
  \providecommand\BibTeX{{%
    \normalfont B\kern-0.5em{\scshape i\kern-0.25em b}\kern-0.8em\TeX}}}

\setcopyright{acmcopyright}
\copyrightyear{2018}
\acmYear{2018}
\acmDOI{10.1145/1122445.1122456}

\acmJournal{TOG}
\acmVolume{37}
\acmNumber{4}
\acmArticle{111}
\acmMonth{8}

\begin{document}

\title{From Cyber Terrorism to Cyber Peacekeeping:
Are we there yet?}

\author{Maria Papathanasaki, Georgios Dimitriou}
\affiliation{%
  \institution{University of Thessaly}
  \city{Lamia}
  \country{Greece}
}

\author{Leandros Maglaras, Ismini Vasileiou}
\email{leandros.maglatas@dmu.ac.uk}
\affiliation{%
  \institution{De Montfort University}
  \city{Leicester}
  \country{UK}}

\author{Helge Janicke}
\affiliation{%
 \institution{Cyber Security Cooperative Research Centre}
 \country{Australia}}

\renewcommand{\shortauthors}{Papathanasaki, et al.}

\begin{abstract}
  In Cyberspace nowadays, there is a burst of information that everyone has access. However, apart from the advantages the Internet offers, it also hides numerous dangers for both people and nations. Cyberspace has a dark side, including terrorism, bullying, and other types of violence. Cyberwarfare is a kind of virtual war that causes the same destruction  that a physical war would also do.  In this article, we discuss what Cyberterrorism is and how it can lead to Cyberwarfare.
\end{abstract}



\maketitle

\section{Introduction}
From the first years of its existence, people undeniably benefit from the Internet. However, only a few comprehend that the e-world is fraught with danger. Unfortunately, Cyberspace has a dark side, including terrorism, bullying, and other types of violence. The dark web is becoming more and more popular among the youth and a portion of criminals \cite{vilic2017dark}. Cyberwarfare is every action in Cyberspace and targets against a country's power or other non-governmental entities (companies, organizations, etc.). Cyberwarfare can cause physical destruction via computers. It is accomplishing military operations using virtual means. This way, countries manage to achieve missions that would require the physical presence of the army. For instance, China regularly cyber attacks Taiwan to weaken the economy of the country. It is essential to note that Cyberwarfare is still a kind of virtual war that causes the same destruction to a state that a physical war would also do. Of course, there is a huge possibility that an e-war can lead to a physical one, causing even more destruction. In summary, we could say Cyberwarfare is the techniques and tactics a country uses virtually and physically, simultaneously or alternately, for an extended period against another state.

\begin{figure}[H]
\centering
\includegraphics[width=0.95\linewidth]{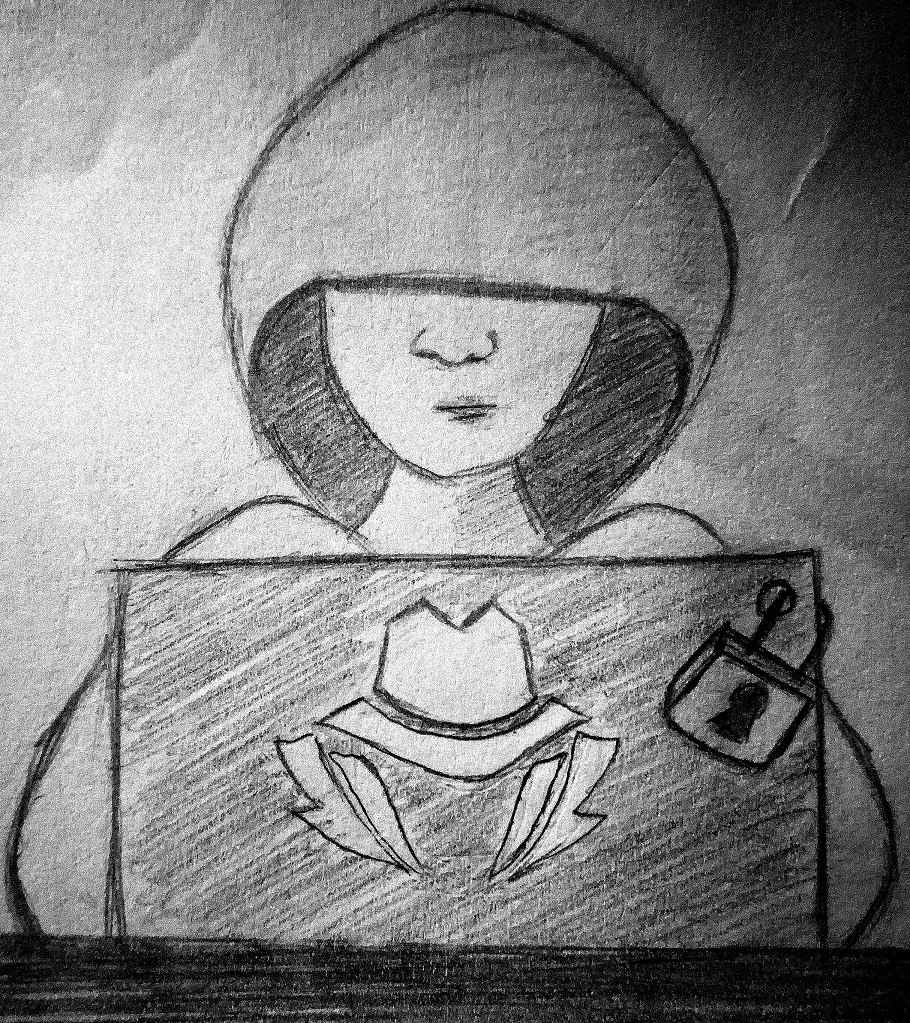}
\caption{Hackers can spread fear}
\label{fig:Fig01}
\end{figure}

\section{Cyberwarfare}
There are many types of warfare that act individually or combined. The most important of them are presented below:

\begin{enumerate}
\item {\bf Espionage: }
Several countries have been accused of spying over others using their secret agencies by recording phone calls in countries like the Bahamas, Afghanistan, Mexico, and others. Whereas some countries resorted to spying on the electronic diplomatic communication channels.
\item	{\bf Sabotage: }
To better understand that type of cyber warfare, we will talk about Stuxnet. In June 2010, a virus named Stuxnet made its appearance in power plants, traffic control systems, and factories worldwide.  Stuxnet appears to be twenty times more complex than the most complex virus until then and has many powers; one of those was the ability to turn up the pressure inside nuclear reactors. It is a weapon made entirely out of code. What makes Stuxnet even scarier is its capability to make everything seem familiar to the engineers. That virus could enter into the systems by security gaps that system creators were unaware of, named zero days. Such gaps can be sold on the black market for $100.000$, and Stuxnet bought twenty of them. The virus's creator targeted a nuclear factory in Iran and managed to shut down a thousand centrifuges in the reactors, leading the Iranian government to suspend the work of all its nuclear facilities without any explanation. Stuxnet infected their nuclear factories, and should nobody notice that, it would lead to a national electricity blackout. What was Iran's response to the attack? They made an open call for hackers to join the Iranian Revolutionary Guard, creating the second-largest online army worldwide. To this day, we do not know for sure who was behind the attack \cite{langner2011ralph}). Nine months after the attack, Stuxnet was redesigned, this time even better, to be able to destroy oil pipelines or power grids, and it is available to anyone to download. Stuxnet is an open-source weapon, free for everyone to play with, including the next person who will use it against a nation.

\item {\bf Denial of service attack: }
Such type of attack aims to make people unable to reach a network resource or make a machine unavailable to users. Attackers often target banks, credit card payment gateways, and other high profile web servers.

\item {\bf Electrical power grid: } 
An electrical power grid is a network for delivering electricity from producers to consumers. A country can infiltrate another country's electrical grid and leave inside any virus that can disrupt the whole system, like an attack that took place in April 2009. Another example is the attack against a nation, causing a power outage for twelve hours. The most recent attack was made against Russia in June 2019, fortunately without consequences because of the Russian government's prompt actions. It is recommended that all countries disconnect the power grid from the Internet and run the net with droop speed control, to avoid such attacks \cite{halpern2015iran}.

\item {\bf Propaganda: }
Cyber propaganda is any type of misinformation, and psychological control of people, using the Internet. According to Jowett and O'Donnell, "propaganda is the deliberate, systematic attempt to shape perceptions, manipulate cognitions, and direct behavior to achieve a response that furthers the desired intent of the propagandist" \cite{jowett2018propaganda}. People who try to propagandize others often use any way possible to brainwash internet users in order to make them serve their purposes. 
\item {\bf	Economic disruption: }
In 2017 a huge attack took place in Ukraine's and U.S.'s National Health Service, using malicious software to make disruptions. Those attacks aim to harm a company's economy, which is also referred to as financial crime \cite{perlroth2017cyberattack}

\item	{\bf Surprise cyber attack: }
Such attacks have as a primary goal to draw attention. The bigger the attention they get, the more successful the attack is. A well-known surprise cyberattack was al-Qaeda's 9/11 attack against the USA, which was broadcasted worldwide, and up to this day, it still remembered.
\end{enumerate}

\subsection{Motivation and Ethics}

What makes Cyberwarfare appealing to more and more countries is a very simple reason. As always, money counts, and in the case of a war, they count even more. Cyberwar costs less than a physical one, and it also offers the opportunity to weaken another nation without risking people's lives. Most of the attacks are politically motivated. For instance, in 2008, hackers attacked CNN. 
Another motivation for causing a cyber attack is the sabotaging Internet itself. Hackers are trying to break into web servers, communication links, businesses, and homes, harming internet service providers, electrical grids, financial networks, etc.
Some attacks are used to produce income for the attacker. Some ransomware can be used by countries to ensure a noticeable profit causing long-term damages to their targets. From the other side of the coin, some organizations are motivated by the need of our time for web safety. For example, Kaspersky Lab examines the issue of Cyberwarfare and tries to raise awareness amongst the internet community. Traditional wars are guided by the Just War Theory (JWT). Several well-defined principles state when a nation is ethically justified to go to war and remain ethical during one. However, are these principles applicable when it comes to cyber-warfare? It is essential to mention that a physical war should be avoided until all other options have been exhausted. On the other side, a Cyberwar is preferable against a physical one, since bloodshed and material damage do not directly occur. People from different nations may not kill each other as it is used to happen in real wars, but a cyberattack against power and food supplies can lead to numerous deaths too. Currently, there are no agreed ethical guidelines for Cyberwarfare. Some researchers have tried to transfer the existing legislature into the cyber world, but warfare ethics are unstable to this day.

\subsection{Opposing Cyberwarfare}

Modern society depends on the Internet more than ever. To ensure our safety, it has been stated that sub-webs should be built since rebuilding the whole Internet is extremely difficult. Sub-webs are going to be equipped with the latest safety protocols. In 2012 a company named Artemis expressed interest in creating such a "secure place". The user had to type .secure at the end of the site address he wanted to visit and was part of the Artemis sub-web. However, Artemis offered safety only to the websites and not the users themselves, exposing them to any kind of malware. Up to date, nobody has managed to make a sub-web safe enough for users.
Another solution is to cut the Internet off from an area that is being attacked. That would be a great solution if it were not for the citizens of that area who would not appreciate the inability to connect to the Internet. An attack that took place in 2012 worldwide almost led the FBI to shut down the connection of the infected area in which more than a million computers existed. To avoid that, they installed two of their own internet servers until they arrested the attackers. All organizations ought to have emergency servers to use in case of an attack.

Last but not least, we have to understand the limits of each country into Cyberspace. There are limits to every state power in Cyberspace and should not be exceeded. Moreover, each government has to reorganize its Cybersecurity and try through a new curriculum to help young people acquire more developed critical thinking. 

\section{Cyberterrorism}

In 1997 Dr. Barry C. Collin proposed the term “Cyberterrorism” for the first time. He described it initially as a premeditated attack on computer systems and data by terrorists. A few years later, all internet terrorism activities were included in the definition of that term. Today, cyber terrorism is considered as every terrorism that uses the Internet as a tool or the network as an attack target \cite{wu2019analysis}. 
According to FBI, cyber terrorism is a “premeditated, politically motivated attack against information, computer systems, computer programs, and data which results in violence against non-combatant targets by sub-national groups or clandestine agents”.  

But is Cyberterrorism worth such fear from us? To this day, no attacks related to physical war have taken place, making Cyberterrorism probably a misnomer. P.W. Singer and Allan Friedman in their book \cite{singerallan}, mention that "Cyberterrorism is like a famous T.V. program named "Shark Week," in which is shown that people every year are more likely to die by an accident involving a toilet, than a shark”. Despite this fact, people are more afraid of dying by a shark than a toilet. Why is that happening? Apparently, we are more afraid of what media, films, or others make us fear. We can easily compare the sharks to Cyberterrorism attacks. People nowadays are terrified of that term as much as other reasons for death. However, as it is mentioned above, Cyberterrorism has not killed or hurt anyone physically yet. What is the reason that makes people afraid of Cyberterrorism? Unfortunately, ignorance is a significant enemy of somebody's composure. Most internet users cannot distinguish the real danger from an insignificant threat. 

Cyberterrorism tries to inflame rebellions in the heart of a nation, destroying its peace. However, Cyberterrorism is not only about national infrastructure sabotaging. For example, even a warning message for a bomb in a public building is considered to be Cyberterrorism. Also, that term refers to acts that hackers do against people in order to spread fear, show off their powers, or even destroy their lives by blackmailing them (See Figure \ref{fig:Fig01}).

It is significant to make a distinction between Cyberterrorism and Cybercrime. Although they are similar terms, they should contend with a different approach by society.  Cyberterrorism, in particular, most of the time, has political motives, but Cybercrime, by contrast, takes place on a more personal level. Certainly, both of the purposes of the terms is to cause harm, but the reasons that lead to that are different. The main goal of Cyberterrorism is to discourage the masses from raising their voice and propagandize them.  

\subsection{TYpes of Cyberterrorism}

Cyberterrorism depending on the way of technology is used to cause harm, can be categorized into six parts, according to Susan Brenner \cite{brenner2006light}:

\subsubsection{Weapons of mass destruction}
\begin{figure}[H]
\centering
\includegraphics[width=0.95\linewidth]{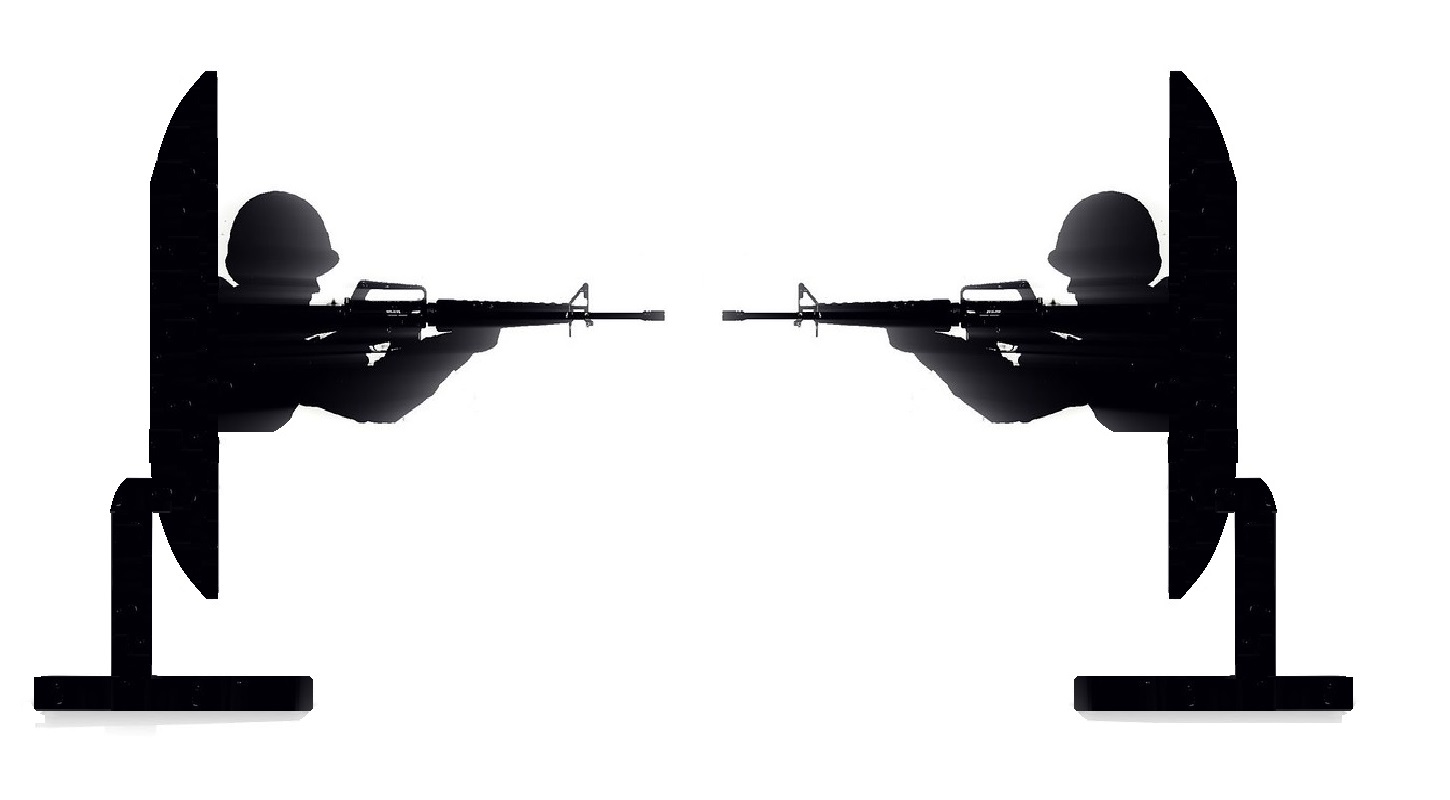}
\caption{Computers can be used as weapons}
\label{fig:Fig0}
\end{figure}

According to S. Brenner, such attacks are not realistic, since computers do not own the power to provoke physical destruction directly to any kind of property. However, they can foment indirect ways that are going to provoke the physical destructions eventually. She also mentions that the Chernobyl disaster in 1986, could have easily been caused by cyber terrorists. For example, they could have broken into the security systems and make the reactors explode. Then they would take advantage of the deaths that happened and turn the people against the government (See Figure \ref{fig:Fig0}).

\subsubsection{Weapons of mass distraction}
To better understand this category, let’s re-inspect the terrorist attacks of the 11th of September 2001 wherein, a terrorist group ‘Al-Qaeda’ (meaning: The Foundation) hijacked four passenger airplanes, setting them out to fly into both world trade center (Twin Towers) and the Pentagon. As a result, the attacks caused 2986 people's death, including the hijackers, and at least $10.000.000.000$ in infrastructure damage. Millions of Americans where watching live the attacks on television, and even more, were trying to reach CNN's official website to get more information about the hijacks. So, what is the factor that would make al-Qaeda terrorism a weapon of mass destruction? Imagine the people who visited the CNN website, instead of seeing the real front page, to see a mirror of that page that shows fake news about other attacks worldwide. That would make people scared even more, and probably it would trigger more terrorist attacks and riots, making computers the real mass destruction weapons.

\subsubsection{Weapons of mass disruption}
Mass disruption weapons aim to make public infrastructure (means of mass transportation, health services, financial institutions, etc.) unreliable to citizens. This time cyber terrorists are trying to cause mental damage to people instead of physical. For example, let's take the "Botnet" malware that managed to turn the Seattle’s Northwest Hospital into chaos. In January 2005, Christopher Maxwell, a twenty-year-old young man from California, infected the hospital network, blocking surgery room doors, moreover doctors’ beeper and intensive care unit’s machines stopped working.

\subsubsection{Cyberwarfare}
It has been clear from the previous chapter what Cyberwarfare is and how it affects modern society. 

\subsubsection{Hybrid warfare}
Hybrid warfare is a kind of Cyberwarfare that apart from uninformed people, others participate too. To be more specific, hybrid warfare combines the army with diplomats, hackers, journalists, and even civilians. All these forces combined, make an extraordinary powerful group that can easily propagandize people. On the 21st of November 2013 in Crimea (Ukraine), civilians rose in protest against the President of Ukraine, because of his denial to sign his country's union with more than four million people were united under one force to fight against the president. Who managed to gather that massive amount of people, though? We cannot be sure who provoked the upheaval, but we know that whoever did this was preparing the whole mission for a long time until the time was right to attack. As Cyberwarfare evolves, more and more hybrid attacks occur, and a new term is born: Unlimited Warfare.

\subsubsection{Unlimited Warfare}
That kind of war is an upcoming way to attack another nation without any barrier. It is said that the first rule of Unlimited Warfare is the absence of rules and limits. Apparently, that "rule" violates any ethical border and leads people to act without any reservations. Unrestricted warfare can be extremely dangerous to modern society due to the lack of respect amongst developed countries. Undeniably the sense of justice and safety is affected during a war. In the case of unlimited warfare, we are talking about a total demolition of what current societies have built.

\subsection{Cyberterrorism and Social Media}

Cyber Terrorism in social media is used for identity theft, online fraud, cyber-attacks, and other reasons. What are the threat categories in social media for cyber terrorism, and how do attackers use social media? As it has been mentioned before, Cyberterrorism is a premeditated electronic attack against civilians to cause harm or spread fear.

Because social media are accessible from people of all ages worldwide, cyberterrorists can easily approach someone and detach useful information, making 90\% of Cyberterrorism to happen through social media. According to Lockheed Martin Corporation, the stages of intrusion kill chain (IKC) are four:

\begin{enumerate}
\item	Information Gathering: collecting information.
\item	Weaponization: Developing malicious code.
\item	Exploitation: Execution of the malicious code.
\item	Installation: Installation of malicious programs.
\end{enumerate}

Usually, cyber terrorists are trying to harm somebody’s reputation, destroy his life, or weaken his mental balance. They also aim to cause more extensive harm, like destroying a company's public profile.

\subsection{Opposing Cyberterrorism}

The Internet is being used more and more every day by people. What measurements should we take to keep ourselves safe, and what do governments ought to do to maintain their stability?
On a personal level, we have to be careful with the information we share. Moreover, we should block any spam accounts we detect and avoid chatting with unknown to us people. We also have to stop using hashtags because they make our profile easier to find. When we face misinformation, we should report them, and double-check the news we read online. It is essential to get informed from reliable sites, and never reading just the title of an article because it could be misleading.

A new growing phenomenon in social media is crowdturfing (crowdsourcing + astroturfing). Crowdturfing uses misleading "posters," which present to their page followers, making them build a negative or positive opinion about a subject by presenting fake views of others. That way of propagandizing can help the spread of cyber-rumors about a person or a company. Crowdturfing puts information integrity and authenticity into risk, so we should have critical thinking before making any decision. For instance, social media can use a pop-up that reminds users about the consequences of misinformation, aiming their ethical logic \cite{parlakkilicc2018cyber}.
It goes without saying that all computers with internet access should have installed updated antivirus software, and all users have to protect their passwords and their sensitive information. 
As Marwan Albahar stated, “if technology can take us to the moon, a breakdown or compromise of the same will ensure that we stay there forever and never return” \cite{seeeuropean2013cybersecurity,hughesnato}. 

\section{Cyber Peacekeeping}
According to United Nations (UN), Cyber Peacekeeping is defined as the “Action undertaken to preserve peace, however fragile, where fighting has been halted and to assist in implementing agreements achieved by the peacemakers”.
Since 1928 UN peacekeeping operations grow more and more. During these years UN has faced many failed processes leading to its radical reformation of its way of acting.
Peacekeeping is an activity that overlaps with a wider set of peace operations. According to the UN, these other activities are \cite{burton2015nato}:

\begin{itemize}
\item	Conflict Prevention  
Early intervention to prevent a dispute escalating.
\item	Peacemaking  
Diplomatic measures aimed at bringing about a ceasefire.
\item	Peace Enforcement 
Restoring peace without the consent of the parties. 
\item	Peace Building 
Laying the foundation for long term peace and preventing relapse into conflict.
\end{itemize}

United Nations follow three peacekeeping principles that are listed below:

\begin{enumerate}

\item	Consent of the parties: 
Peacekeeping operations are only deployed with the consent of the conflicting parties. Without the consent of all parties, the operation risks becoming involved in the conflict.

\item	Impartiality: 
Operations maintain peace without favoring any of the involved parties. An operation must be seen as impartial to remain legitimate. 

\item	Non-use of force, except in self-defense and defense of the mandate: 
The use of force should be a measure of last resort. The force could be used against those who were determined to undermine the peace process.
\end{enumerate}

Cyberwar is considered to be the latest form of war, and it should be treated like every other type of war. That fact leads us to the conclusion that cyberwar could potentially threaten the peace in a country. Preventing a war is preferable than dealing with its aftermath, so it is essential to take every measurement possible to prevent such actions. 

\subsection{Goals and principles}

In our daily lives, peacekeeping is maintained through a set of rules that everyone must obey. These regulations are set by the state and intend to adjust the smooth function of the society. Cyberspace is our society in a small scale, which also needs rules to be kept safe and accessible to everyone.
Human rights, democracy, safety and freedom are only a few of the countless fundamental rights that ought to be protected in Cyberspace. Nowadays, legislature is handling more and more human protection cases and it has become stricter against the ones who offend against it.
According to the European Union Cybersecurity Strategy, Cybersecurity is governed by a principle guide that should be followed internationally \cite{seeeuropean2013cybersecurity}.

\begin{itemize}
\item	Cyberworld follows the same rules as the physical one
Every rule that ensures the safety of the physical world, also applies to the digital society. 
\item	Unlimited access to everyone
The Internet is a huge world, containing everything the physical one has to offer, but in a virtual way. In contrast to the real world, the Internet is accessible to everybody, and offers unlimited information for free. That means that safety is essential in such an extremely large space.  
\item	Protection of users 
Someone’s safety can only be assured in a secure place. According to European Union Charter of Fundamental Rights, all users' data are processed, ensuring that everyone's sensitive personal information remains private.
\item	Security maintenance, is everyone’s responsibility
The Internet is accessed daily by citizens, organizations, public and private authorities. All of the above are partially responsible for their safety, and Cybersecurity is strengthened every time individuals take measurements about their own security.  
\item	Safety control distribution
The size of Cyberspace requires control by many entities. These days, our safety on web rests on the shoulders of non-governmental / governmental organizations. Those entities act according to protocols that EU has defined, and standards that are adapted to the needs of the time.
\end{itemize}

\subsection{Implementation of Cyber Peacekeeping}

Since ancient times, people are trying to find ways to avoid or stop wars. In ancient Greece, during the Olympic Games, people agreed to a truce. Throughout wars, they would dedicate some days to bury the dead soldiers or pray to Gods. All these contradict the fact that at the same time they would make deadlier weapons to destroy races easier. Nowadays a huge concern rises. Perhaps the deadliest weapon of all times has been created; Cyberspace.

It may sound extreme to say that Cyberspace is deadly; however it is possible to lead to a world war faster than a physical one. On the other side, it could also be the way to avoid such wars. 
The Internet is growing stronger every day, and maintaining peace in it is extremely challenging. Peacekeeping seems to be inevitable considering the dangers that lurk, and it is the only way to ensure the safety and peace among the users, who could be individuals or nations.

United Nations often interfere when asked, to offer help to states that are in conflict. For example two countries that wish to stop an imminent war, can turn to UN for assistance. UN often retreat armed soldiers to a defined boundary and use diplomacy to avoid making the situation worse. Bringing the conflict to an end means that Cyberwarfare also stops, leading both countries to an agreement. The complexity of the UN's procedure depends mainly on the states' cooperation and the harm that every country has caused to the other \cite{robinson2019developing}.

To conclude UN peacekeepers Observe, Monitor and Report (OMR) the cases of Cyberwar, and close the safety gaps of Cyberspace, making it safer over time \cite{robinson2018introduction}.

\section{Cyber Defence}

Cyber defence is a computer network defense mechanism which includes response to actions and critical infrastructure protection and information assurance for organizations, government entities and other possible networks. Cyber defense focuses on preventing, detecting and providing timely responses to attacks or threats so that no infrastructure or information is tampered with. With the growth in volume as well as complexity of cyber attacks, cyber defense is essential for most entities in order to protect sensitive information as well as to safeguard assets.

\subsection{NATO’S Cyber Defence}

North Atlantic Treaty Organization (NATO) was created on 4 April 1949 and is an alliance between 30 European and North American Countries. NATO assures these countries' safety against others' attacks and aims to tighten the bonds among the member states.
Since the day NATO was formed, its priority is to defend its part in Cyberspace. To ensure that NATO stays safe, two tools, which are presented below, are combined with monitoring and dealing in real-time all threats:
\begin{enumerate}
\item	The National Criminal Intelligence Resource Center (NCIRC)  NCIRC is based in the U.S. Its objective is all forms of crime prevention and the criminal justice system's improvement.

\item	Code-division multiple access (CDMA)
CDMA is a channel access method applied in most radio communication technologies, allowing users to share a bandwidth of frequencies without interference \cite{torrieri2005principles}.
\end{enumerate}

It should be noted that NATO itself does not have offensive cyber attack capabilities, and up to this day, any offensive act is utterly upon the individual member states.
Moreover, because of the complexity of counterattacking cyber attackers, NATO aims to make them feel that they are unable to achieve a successful attack, avoiding that way many offensive operations before they start.
NATO's power originates from the cooperation of the individual states and state-sponsored hackers. It also approaches Cybersecurity diplomatically and peacefully, trying to develop trust between NATO-states and other countries \cite{burton2015nato}.

\subsection{Cyber Self-defence}

Nowadays, all individual systems that constitute a state, such as military, financial, and health systems, are connected so tightly that a massive attack could cause great harm \cite{waxman2013self}.  Until a few years ago, most kinds of attacks resulted from pranks by small groups of people. However, today, these attacks are under a well-planned procedure that exceeds the prank limits. Instead of the Cyberworld's safety valves, every simple user, company, or state ought to take extra measurements for their safety and privacy protection, following the four factors explained below.

\begin{enumerate}
\item	Culture awareness
Individuals are important to follow simple rules such as creating strong passwords, installing antivirus software, and publicly sharing sensitive information \cite{vasileiou2019cybersecurity}.

\item	Active \& passive defense
Having up to date antivirus, two-factor authentication to all devices and accounts and an active firewall are only few of the many steps that someone could do to prevent attacks.

\item	Breach readiness
It goes without saying that everyone has to keep backups in case of an attack that might cause data loss.

\item	High-value data content management
It is highly recommended that no sensitive and personal data are shared online. Especially any personally identifying data should be avoided to be published under all circumstances. 
\end{enumerate}

By hiring specialized staff to recognize and evaluate the incoming threats, companies will prevent attacks before they cause harm. Staying cyber safe is not hard to achieve, and it depends on us to a great extend.

\subsection{Cyber Defence Technology}

\subsubsection{Vulnerability Assessment(VA) and Penetration Tests (PT)}

Securing systems requires that we find any vulnerabilities before attackers do. That procedure, which is described as Vulnerability Assessment (VA), provides cyber defense at a great extend. At VA all software, networks and systems are checked for weaknesses that leave room to attackers to attack undisturbed. The vulnerabilities that a system may have are listed below.

\begin{itemize}
\item	Authentication Vulnerabilities
\item	Configuration Weakness Vulnerabilities
\item	Access Control Vulnerabilities
\item	Exception Handling Vulnerabilities
\item	Boundary Condition Vulnerabilities
\item	Input Validation Vulnerabilities
\end{itemize}

After a possible vulnerability is discovered, Penetration Tests (PT) take place to investigate the safety gap more. All weaknesses are discovered through various techniques that include both automated and manual testing (software/knowledge-experience). Attackers run VA at the victim using the same techniques and get a list of all possible vulnerabilities. Fortunately, if the victim manages to ensure the systems' safety before the attack, a successful Vulnerability Assessment - Penetration Test (VAPT) will have taken place, and nobody will penetrate them \cite{waxman2013self}.

In conclusion, VAPT tests are recommended to take place in all systems to ensure their maximum cyber safety. Also, more VAPT software and techniques should be created, that will improve the existing ones.

\subsubsection{AI}
There is no doubt that Artificial intelligence (AI) plays an important role in the fight against the cyber crime. First of all, AI can be used in creating intelligent machines, or in solving difficult problems such as in cyber defense \cite{maglaras2014ocsvm, tyugu2011artificial}.

For example, neural networks are popular in cyber defense, because of the high speed neurons learn and solve problems. The most used AI tool however, are the expert systems, that are used for decision support. Expert systems subserve the decision making in many domains including cyberspace. They contain an extensible database of scientific knowledge, and they can support solving of a great range of problems. The ability of the expert systems to solve problems depends on the quality of the knowledge they contain \cite{rosenblatt1957perceptron}.

An ideal way to cope with DDoS attacks is the intelligent agents software components. Intelligent agents are able to make decisions and act accordingly, depending on the circumstances.  Due to some legal and commercial problems they are not widely used, but as soon as these obstacles are overcomed, intelligent agents should be used to implement a powerful infrastructure against DDoS attacks \cite{kotenko2007multi,kotenko2010agent}. 

Artificial intelligence is constantly developing ways to improve the existing systems. This is possible through enriching current knowledge databases with new information, or by reorganizing them. The huge amount of data that we have at our disposal is often hard to manipulate, so data mining was developed. Data mining was born through unsupervised type of machine learning, and it is widely used in cyber defense, because of the safety gaps that occur from that uncontrollable amount of data. In order to help the systems learn easier and faster, new algorithms and learning methods are developed constantly, making threat detection systems more productive.

Last but not least, a common technique used in problem solving and decision support, is constraint satisfaction. In that process, the problem requires its solution within some limitations also known as constraints. Depending on the constraints, the techniques that are used vary. A typical example of constraint satisfaction is the game Sudoku. In that game the player is asked to fill a table 9x9 with numbers from 1 to 9 when a part of the table is pre-filled. The constraints in that problem solving game, is that the lines, rows and blocks of the table, must not contain the same number more than once. The solution of the problem is the one that satisfies all constraints \cite{tyugu2007algorithms}.  

As it was mentioned before, AI methods are used more and more on Cyber Defense, and in the future it seems that they will be an indefeasible part of it. The power of cyber attackers is growing rapidly, however the techniques that were mentioned above, have the power to hinder their attacks if used properly.

\section{Conclusions}

Research into conducting and understanding cyber warfare and cyber terrorism is extensive and wide-ranging \cite{robinson2019developing}, yet research into restoring peace after cyber warfare has recently been addressed \cite{ayres2016cyberterrorism}. Attribution of cyber-attacks is an open issue \cite{cook2016attribution} and the correct norms and procedures are yet to be discovered. 
Some security solutions  specifically for Critical Infrastructures must be put forward in the National or International level. Undertaking efficient cyber security risk assessments, maturity assessments \cite{drivas2020nis} and implementing mitigation plans ae important steps for securing CNIs \cite{cook2017industrial}. 

There is still much work to be done on Cyber Peacekeeping and Cyber Defense, but soon enough, they will be necessary to avoid cyber attacks. Maintaining peace will be hard to accomplish in the near future; however, more and more organizations apart from NATO and UN will be available to defend peace in Cyberspace.

\bibliographystyle{ACM-Reference-Format}
\bibliography{sample-base}

\end{document}